\documentstyle[11pt,aaspp4]{article}
\def\pmode{{\it p}\/-mode}
\def\pmodes{{\it p}\/-modes}

\def\dd{{\rm d}}

\def\Ai{{\rm Ai}}
\def\Bi{{\rm Bi}}

\begin{document}

\title{Peaks and Troughs in
Helioseismology: The Power Spectrum of Solar Oscillations}
\author{C. S. Rosenthal}
\affil{High Altitude Observatory, National Center for Atmospheric Research,
P.O. Box 3000, Boulder, Colorado, 80307-3000, USA}
\authoremail{rosentha@hao.ucar.edu}

\begin{abstract}

I present a matched-wave asymptotic analysis of the driving of solar
oscillations by a general localised source. The analysis provides a simple
mathematical description of the asymmetric peaks in the power spectrum 
in terms of the relative locations of eigenmodes and troughs in the spectral
response. It is suggested that the difference in measured phase function
between the modes and the troughs in the spectrum will provide a key
diagnostic of the source of the oscillations. I also suggest a form for the
asymmetric line profiles to be used in the fitting of solar power spectra.

Finally I present a comparison between the numerical and asymptotic descriptions 
of the oscillations. The numerical results bear out the qualitative features
suggested by the asymptotic analysis but suggest that numerical calculations
of the locations of the troughs will be necessary for a quantitative
comparison with the observations.

\end{abstract}


\section{Introduction}

Helioseismology is primarily concerned with the peaks in the solar oscillation
spectrum which are identified with acoustic eigenmodes of the Sun. However
the whole spectrum, including the variation between the mode peaks, contains
useful information about the Sun's structure and dynamics. Moreover, 
the asymmetries in the \pmode~lines which were first identified by \markcite{D93} Duvall 
et al. (1993) can bias the fitting of profiles to the peaks
and hence introduce errors in the determination of the frequencies. The 
most recent space-based spectra from SOHO-MDI \markcite{Netal98} (Nigam et al. 1998)
have confirmed the existence of a strong asymmetry. Thus the study of
\pmode~line asymmetries is prompted both by the need to improve on traditional
data analysis techniques and by the hope of obtaining new scientific
information from them.

The occurrence of highly asymmetric line profiles is a common feature of 
models in which the oscillations are driven by a localised source
\markcite{G92,G93,G95}(Gabriel
1992, 1993, 1995). \markcite{AK96}Abrams and Kumar (1996) explored the nature of
the asymmetry in a simple model using analytical techniques and then applied these
ideas to the interpretation of numerical calculations of the line asymmetry
in a real solar model, including the variation of the asymmetry 
with frequency and mode degree and estimates of the error introduced into the
determination of the frequencies by the neglect of the line asymmetry.

\markcite{RB97}Rast \& Bogdan (1997) emphasised the sensitivity of the asymmetry to
the depth and type of the source. In particular they introduced the approach
that the observed asymmetry can best be approached through understanding the
relative locations of the peaks and troughs in the spectrum, where the
location of the troughs can be determined (in some circumstances) by solving
an eigenproblem for a solar model truncated from above at the source depth. 
Although the calculations presented here will result in an estimate for 
the whole spectrum, I will follow the example of Rast \& Bogdan in concentrating
particularly on the relative locations of the peaks and troughs as a key
diagnostic indicator of the source of excitation.

\markcite{RV95}Roxburgh \& Vorontsov (1995) used a form of asymptotic analysis to model
the power spectrum. The analysis presented here differs from theirs in that
it uses a matched wave technique in which the matching across the turning 
point(s) is treated explicitly. Moreover, I consider a general form of
the source term and obtain an explicit expression for the asymptotic
line profile.
It is found that the power spectrum can be represented in a simple
form even for a general source, and this should guide the analysis 
and interpretation of the peaks and troughs in the solar \pmode~spectrum.

I begin in \S\ref{illus} with a very simple illustrative model which nevertheless
captures some important aspects of the dynamics. In \S\ref{strat} I 
show how one may carry out a matched-wave asymptotic analysis for
a stratified layer with a source containing both monopole and dipole terms.
I am predominantly interested in the lowest frequency modes, for which the
asymmetry is most easily observed. However, asymptotic results for intermediate
and high-frequency oscillations are also presented. In \S\ref{results} I 
evaluate the resulting asymptotic expressions for a simple model and compare them
with a direct numerical calculation. 

\section{An Illustrative Model}
\label{illus}

Figure \ref{fig1} shows a simple model of a driven oscillator consisting of a
uniform string characterised by a soundspeed $c$, clamped at its ends at $z=0,L$ 
and driven by a localised source
at $z=z_0\ll L$. The displacement, $\Psi$, of the string satisfies the equation
\begin{equation}
{\dd^2\Psi\over\dd z^2}+k^2\Psi = a_0\delta(z-z_0)+a_1\delta^\prime(z-z_0)
\label{eq1}
\end{equation}
where the wavenumber $k=\omega/c$ and the amplitudes of the monopole and dipole
source terms, $a_0$ and $a_1$, are, in general, arbitrary complex functions of
the frequency $\omega$. 

Although such a model is almost trivial to analyse, its solution will be found to 
illustrate many of the properties of the more general problem with which we are
primarily concerned. We can obtain jump conditions at $z=z_0$ by integrating the
equation of motion from $z=z_0-\epsilon$ to $z=z_0+\epsilon$ and by multiplying the
equation by $z$ and then performing the same integral:
\begin{equation}
\left.{\dd\Psi\over\dd z}\right|_{z_0+\epsilon}-
\left.{\dd\Psi\over\dd z}\right|_{z_0-\epsilon}=a_0
\end{equation}
and
\begin{equation}
\Psi_{z_0+\epsilon}-\Psi_{z_0-\epsilon}=a_1.
\end{equation}
Evidently a dipole source is inconsistent with a physically realisable string, which
must be everywhere continuous, but it is a useful idealisation;
equivalent to a pair of oppositely-signed monopole sources whose separation is
much less than any other length-scale in the problem.
The solution to equation (\ref{eq1}) is then 
\begin{equation}
\Psi=\cases{A\sin kz, &$z<z_0$,\cr
            B\sin k(z-L) &$z>z_0$,\cr}            
\end{equation}
where
\begin{equation}
A={1\over k\sin kL}\left[ -ka_1\cos k(z_0-L) +a_0\sin k(z_0-L)\right]
\end{equation}
and
\begin{equation}
B={1\over k\sin kL}\left[ -ka_1\cos kz_0 +a_0\sin kz_0\right].
\end{equation}
It is immediately evident that the amplitudes both diverge for $kL=n\pi$, {\it i.e.}\/ at
an eigenmode of the whole cavity.

Now in the Sun, the driving occurs close to the surface but below the region where the
observations are made so, by analogy, we define the measured power as $|A|^2$. However the
total energy in the string is proportional to $z_0|A|^2+(L-z_0)|B|^2$. For
convenience we define a quantity 
\begin{equation}
\mu={a_0+a_1\over a_0-a_1}.
\end{equation}
Figure \ref{fig2} shows plots of the power and energy spectra for several choices of 
$\mu$ and $z_0$. The following properties of the solutions are clearly evident:
\begin{description}
\item[(i)] The degree of asymmetry in the power spectrum increases as the source is
           moved closer to the upper boundary.
\item[(ii)]The peaks in the energy spectrum are much more symmetric than those in the
           power spectrum. Thus, in a sense, the zeros in the power spectrum are
           illusory, or at least misleading, since there are no zeros in the energy spectrum excited
           by the source. 
\item[(iii)]For
             the lowest frequency modes, and for a monopolar source, the 
             troughs lie above the frequency of the corresponding modes (a 
             negative asymmetry in the language of Abrams \& Kumar). 
\item[(iv)]For a monopole source, the asymmetry decreases as the frequency is
           increased and eventually changes sign.
\item[(v)]A purely dipolar source results in peaks with the opposite sense of asymmetry to
          that given by a monopole source. 
\item[(vi)]For a dipole source, the asymmetry increases as the frequency is increased
           and eventually reverses sign.
\end{description}

As we shall show below, all these properties are also present in the power spectrum for
a stratified medium with the exception of property (vi). Instead, for the stratified 
case, a purely dipolar source located in the evanescent region gives an asymmetry which 
decreases in magnitude with increasing frequency. 

\section{Asymptotic Analysis for a Stratified Atmosphere}
\label{strat}

We restrict ourselves to consideration of the oscillations of a plane-parallel
stratified layer and ignore the effects of curvature. Oscillations 
of such a region with frequency $\omega$ and horizontal wave number $k$ satisfy the equation
\begin{equation}
{\dd^2\Psi\over \dd z^2}+K^2\Psi=A_{\omega,k}(z)
\label{general_eqn}
\end{equation}
where the independent variable $z$ increases in a downward direction and the 
dependent
variable $\Psi=\rho^{1/2}c^2\nabla\cdot{\bf u}$, where $\rho$ and $c$ are 
the equilibrium density and sound-speed respectively and ${\bf u}$ is the
velocity field of the perturbation. The form of the source term, $A_{\omega,k}(z)$, is discussed
in detail by Gabriel (1993).
The local vertical wavenumber, $K$, satisfies
\begin{equation}
K^2={1\over c^2}\left[ \omega^2-\omega_c^2-c^2k^2\left(1-{N^2\over\omega^2}\right)\right]
\end{equation}
where
\begin{equation}
\omega_c^2={c^2\over 4H^2}\left(1+2{\dd H\over\dd z}\right)\hbox{~~;~~}
H=\left({\dd\ln\rho\over\dd z}\right)^{-1}
\end{equation}
and
\begin{equation}
N^2=g\left( {\dd\ln\rho\over\dd z}- {1\over \Gamma_1}{\dd\ln p\over\dd z}\right)
\end{equation}
and where $p$ and $g$ are, respectively, the equilibrium pressure and the
acceleration due to gravity and $\Gamma_1=({\partial \ln p /\partial\ln\rho})_s$, the 
derivative being at constant entropy.

\subsection{Driving in the Evanescent Region}
\label{case1}

For a solar-like stratification, and for frequencies
less than the photospheric cutoff at about 5.3mHz, there is a cavity in which 
$K^2>0$,
bounded by turning points at which $K^2$ becomes negative. The lower turning point is
dictated by the rapid increase of $c$ with depth, while the upper turning point is
dictated by the rapid increase of $\omega_c$ near the solar photosphere. 
Modes
close to the cutoff have an upper turning point very close to the surface while
lower frequency modes are reflected somewhat more deeply. 

The precise form of the source terms in equation (\ref{general_eqn}) cannot be deduced
without a complete theory of solar convection. However, it is clear that the source
must be strongly peaked in the turbulent super-adiabatic boundary layer at the
top of the convection zone, a result also evident from the observationally-determined
excitation rate \markcite{GMK94}(Goldreich et al. 1994). In this section we will restrict 
our analysis to the
case of low-frequency oscillations for which this source lies in the evanescent region
above the upper turning point. 

For simplicity we will consider the case of a source localised at a single height:
\begin{equation}
A_{\omega,k}(z)=a_0|K(z_0)|^{1\over 2}\delta(z-z_0)+a_1|K(z_0)|^{-{1\over 2}}\delta^\prime(z-z_0)
\label{local}
\end{equation}
where the factors $|K(z_0)|^{\pm {1\over 2}}$ have been introduced purely for convenience
(since $a_0$ and $a_1$ may, in general, be functions of frequency and wavenumber).
The solution for a more general source term will be given in \S\ref{App}.
In the following analysis I shall refer to the terms involving $a_0$ and $a_1$ as
monopole and dipole terms respectively. This use of the terms ``monopole'' and
``dipole'' is quite distinct from that employed by \markcite{GK90}Goldreich \& Kumar
(1990) who define monopole terms as those sources involving a volume change, dipole
terms as those arising from a momentum source, and quadropole terms as those arising
from internal stresses. This latter definition is clearly the better physical 
description of the actual excitation mechanism. Moreover, as Goldreich \& Kumar 
point out, a definition of the multipole expansion solely in terms of the
number of derivatives appearing in the source terms is ambiguous unless one
also specifies ones preferred choice of independent wave variable. However,
given a particular choice of wave variable (and the variable $\Psi$ is a
natural choice) the form of the power spectrum is indeed
determined by the number of derivatives (or rather the number of sign
changes) appearing in the source terms. The precise relationship between
the two descriptions of the source terms depends on the detailed properties
of the turbulent convection itself. In the future we hope to expand on
the numerical work of \markcite{SN91} Stein \& Nordlund (1991) to obtain a fully
consistent picture of the excitation process. However, for the present 
purposes, the terms ``monopole'' and ``dipole'' will be used to refer
to terms involving $a_0$ and $a_1$ as above, with no particular assumption
being made as to the physical nature of the driving mechanism.

The geometry of the problem is indicated in figure (\ref{fig4}). The propagating region
(Region II) is bounded by the lower and upper turning points $z_t$ and $Z_t$ respectively,
and the source at $z_0$ divides the upper evanescent region in two. We therefore proceed
by writing down approximate solutions to equation (\ref{general_eqn}) valid in each of the
four regions and using matching conditions at the boundaries between the regions to
fix the arbitrary constants appearing in the solution.

I shall use the (lowest order) WKB solutions, which are valid provided the scale of
variation of $K^2$ is long compared to the vertical wavelength, {\it i.e.}
\begin{equation}
\left|{\dd |K|\over\dd z}\right|\ll|K^2|.
\end{equation} 
This condition will certainly break down close to the
turning points where $K=0$. The lowest-order WKB solutions for the geometry of figure \ref{fig4}
are
\begin{eqnarray}
\rm{I:~}\Psi\sim& {1\over \left|K^2\right|^{1\over 4}} A
                   \exp\left[ -\int_{z_t}^z \left| K^2(z')\right|^{1\over 2}\dd z'\right],
                   & z\gg z_t \cr
\rm{II:~}\Psi\sim&{1\over K^{1\over 2}}\left\{
         B\exp\left[ i\int_{Z_t}^z K(z')\dd z'\right]+C\exp\left[-i\int_{Z_t}^z K(z')\dd
         z'\right] \right\},
         & z_t\gg z\gg Z_t\cr  
\rm{III:~}\Psi\sim &{1\over \left|K^2\right|^{1\over 4}} \left\{   
          D\exp\left[ -\int_{z_0}^z \left| K^2(z')\right|^{1\over 2}\dd z'\right]  
         +E\exp\left[ \int_{z_0}^z \left| K^2(z')\right|^{1\over 2}\dd z'\right] \right\},
         &Z_t\gg z > z_0\cr
\rm{IV:~}\Psi\sim&{1\over \left|K^2\right|^{1\over 4}} F
                   \exp\left[ -\int_{z_0}^z \left| K^2(z')\right|^{1\over 2}\dd z'\right],
                   &z<z_0
\label{WKB1}                   
\end{eqnarray}
where $A-F$ are (as yet undetermined) constants.
A requirement for the validity of this analysis is clearly that $z_0\ll Z_t$, which
is to say that the source must lie in a region in which the waves are highly evanescent -
the low-frequency limit.

The jump conditions at $z=z_0$ are determined exactly as in \S \ref{illus}:
\begin{equation}
\left.{\dd\Psi\over\dd z}\right|_{z_0+\epsilon}-
\left.{\dd\Psi\over\dd z}\right|_{z_0-\epsilon}=a_0|K(z_0)|^{1\over 2}
\label{jump1}
\end{equation}
and
\begin{equation}
\Psi_{z_0+\epsilon}-\Psi_{z_0-\epsilon}=a_1|K(z_0)|^{-{1\over 2}},
\label{jump2}
\end{equation}
giving 
\begin{equation}
-D-E+F=a_1
\label{coeff1}
\end{equation}
and
\begin{equation}
D-E+F=-a_0
\label{coeff2}
\end{equation}
when derivatives of $K^2$ are neglected.

The determination of jump conditions at the turning points is essentially standard analysis.
In the region of the upper turning point, $K^2$ is assumed to vary linearly so that
\begin{equation}
K^2=\beta(z-Z_t)
\label{beta}
\end{equation}
and the solution can be written
\begin{equation}
\Psi\approx\alpha_1\Ai(\beta^{1\over 3}(Z_t-z))+\alpha_2\Bi(\beta^{1\over 3}(Z_t-z))
\label{Airy}
\end{equation}
where $\Ai$ and $\Bi$ are Airy functions and $\alpha_1$ and $\alpha_2$ are constants
whose linear relationship to $B,C,D$ and $E$ is to be determined. We can substitute
the expansion (\ref{beta}) into the WKB solution (\ref{WKB1}) to obtain
\begin{equation}
\label{e1}
\Psi\approx\beta^{-{1\over 4}}(z-Z_t)^{-{1\over 4}}
           \left\{ Be^{{2\over 3}i \beta^{1\over 2}(z-Z_t)^{3\over 2}} +
                   Ce^{-{2\over 3}i \beta^{1\over 2}(z-Z_t)^{3\over 2} } \right\}
\end{equation}
in the region $z>Z_t$ and
\begin{equation}
\label{e2}
\Psi\approx\beta^{-{1\over 4}}(Z_t-z)^{-{1\over 4}}
           \left\{ De^{-\Delta}e^{{2\over 3} \beta^{1\over 2}(Z_t-z)^{3\over 2}} +
                   Ee^{\Delta}e^{-{2\over 3} \beta^{1\over 2}(Z_t-z)^{3\over 2} } \right\}
\end{equation}
in the region $z<Z_t$. Here we have introduced
\begin{equation}
\Delta\equiv\int_{z_0}^{Z_t} |K^2|^{1\over2} \dd z.
\label{Delta}
\end{equation}
The Airy function expansion (eq. \ref{Airy}) can be expanded asymptotically for large
negative or positive arguments to give
\begin{equation}
\label{e3}
\Psi\approx\pi^{-{1\over 2}}\beta^{-{1\over 12}}(z-Z_t)^{-{1\over 4}}
\left\{ \alpha_1\sin\left[ {2\over 3}\beta^{1\over 2}(z-Z_t)^{3\over 2}+{\pi\over 4}\right]
+ \alpha_2\cos\left[ {2\over 3}\beta^{1\over 2}(z-Z_t)^{3\over 2}+{\pi\over 4}\right]
\right\}
\quad\quad z\gg Z_t
\end{equation}
and
\begin{equation}
\label{e4}
\Psi\approx\pi^{-{1\over 2}}\beta^{-{1\over 12}}(Z_t-z)^{-{1\over 4}}
\left\{ {1\over 2}\alpha_1 e^{- {2\over 3}\beta^{1\over 2}(Z_t-z)^{3\over 2} }
+ \alpha_2 e^{ {2\over 3}\beta^{1\over 2}(Z_t-z)^{3\over 2} }
\right\}
\quad\quad z\ll Z_t.
\end{equation}

It is now straightforward to match the coefficients of exponentials in the overlap
region between equations (\ref{e1}) and (\ref{e3}) and between (\ref{e2}) and 
(\ref{e4}) to obtain four equations from which one may eliminate $\alpha_1$
and $\alpha_2$ to obtain
\begin{equation}
-Be^{-{i\pi\over 4}}+Ce^{{i\pi\over 4}}=2iEe^\Delta
\label{coeff3}
\end{equation}
and
\begin{equation}
Be^{-{i\pi\over 4}}+Ce^{{i\pi\over 4}}=De^{-\Delta}.
\label{coeff4}
\end{equation}
Similar analysis at the lower turning point gives
\begin{equation}
Be^{i\phi}e^{{i\pi\over 4}}+Ce^{-i\phi}e^{{-{i\pi\over 4}}}=0
\label{coeff5}
\end{equation}
where
\begin{equation}
\phi\equiv\int_{Z_t}^{z_t}K\dd z.
\label{phi}
\end{equation}
The five equations (\ref{coeff1}, \ref{coeff2}, \ref{coeff3}, \ref{coeff4} and \ref{coeff5})
determine the five amplitude coefficients $B$-$F$. (A second condition at the
lower turning point can be used to determine $A$ if it is required.)

In the solar case, the oscillations are measured above both the upper turning point and
the source so the measured amplitude spectrum will be the quantity $F$ (multiplied by some
slowly varying function of frequency to convert the spectrum of $\Psi$ to a doppler
velocity spectrum). It is straightforward to solve the equations to obtain 
\begin{equation}
F={1\over 2}(-a_0-a_1)+{1\over 4}(-a_0+a_1)e^{-2\Delta}\tan\phi
\label{spec1}
\end{equation}
which is the low-frequency asymptotic amplitude spectrum.

\subsubsection{Properties of the Power Spectrum}

The spectrum defined by equation (\ref{spec1}) diverges at frequencies corresponding
to eigenmodes of the cavity, {\it i.e.}\/
\begin{equation}
\int_{Z_t}^{z_t}K\dd z=\left(n-{1\over 2}\right)\pi
\label{modes1}
\end{equation}
independent of the form and location of the driving.
However the minima of $|F|^2$ depend on the values of
$z_0$,
$a_0$ and $a_1$. Thus, as we would expect, the spectral response away from the eigenmodes
contains information about both the cavity and the source. If we consider cases where the
ratio $(a_1-a_0)/(a_1+a_0)$ is real then the minima will, in fact, be zeros (nulls). 
We are particularly interested
in the low-frequency/large-$\Delta$ case, for which it is evident that the zeros must
lie close to the eigenmodes, {\it i.e.}\/ for which the peaks in the power spectrum must
be most asymmetric. In this case it is easy to show that the zeros are given by
\begin{equation}
\int_{Z_t}^{z_t}K\dd z\approx\left(n-{1\over 2}\right)\pi+2e^{-2\Delta}{ a_0+a_1\over a_0-a_1}.
\label{zeros1}
\end{equation}
Now the integral in equation (\ref{phi}) is known
\markcite{CDPH}(Christensen-Dalsgaard \& P\'erez Hern\'andez 1992) to have 
the property
\begin{equation}
\int_{Z_t}^{z_t}K\dd z\approx\omega F_D\left({\omega\over k}\right)+I(\omega)
\end{equation}
where the function $F_D(\omega/k)$ is the Duvall function and $I(\omega)$ is a
term whose properties depend predominantly on the structure of the surface layers.
Furthermore, it is clear that $\Delta$ is almost independent of $k$ since it involves
the evaluation of $K$ only in the evanescent region where $\omega\gg ck$. Thus
both the dispersion relation [eq. \ref{modes1}] and the equation for the nulls [eq.
\ref{zeros1}] can be written in the form
\begin{equation}
F_D\left({\omega\over k}\right)=\pi{ n+\alpha(\omega)\over\omega }
\end{equation}
where the Duvall function is the same in each case but the 
phase function $\alpha(\omega)$ for the troughs differs 
from that from the modes. In the limit of large $\Delta$ we have
\begin{equation}
\alpha_t-\alpha_m= {2\over\pi}e^{-2\Delta}{ a_0+a_1\over a_0-a_1}
\label{dalpha}
\end{equation}
where $\alpha_t$ and $\alpha_m$ are the phase functions specifying the positions
of the troughs and modes respectively.
The sign of the difference
depends on the type of the source. For a pure monopole source, $\alpha_t-\alpha_m$
is positive, corresponding to troughs in the
power spectrum lying above the corresponding peaks, with the situation being 
reversed for a pure dipole source. In \S \ref{results} I will show results
for a simple model which illustrate these properties of the power spectrum.

The special case $a_0=a_1$ corresponds to a solution with $A=B=C=D=E=0$. In this
case there is no wave energy below the source and so the spectrum contains
no signal of the presence of an acoustic cavity. Thus the spectrum of $F$ is flat.

\subsubsection{The General Solution}
\label{App}

If we consider the solution to the local-driving problem [eq. \ref{local}] with
$a_1$ set
to zero then we immediately recover the Green's Function for the general
case of equation (\ref{general_eqn}). Thus, if the source term is located entirely
within the evanescent region one can simply write down the solution in the region
above
the source:
\begin{equation}
\Psi\sim { \exp\left[ \int _{Z_t}^z \left| K^2(z')\right|^{1\over 2}\dd z' \right]
\over
4\left| K^2(z)\right|^{1\over 4} }
\int A_{\omega,k}(z_0)\left\{ 2e^{\Delta(z_0)} + e^{-\Delta(z_0)}\tan\phi\right\} 
\left| K(z_0)\right|^{-{1\over 2}}\dd z_0
\label{general_soln}
\end{equation}
where $\Delta$ and $\phi$ are defined by equations (\ref{Delta}) and (\ref{phi}) 
respectively. Note that $\phi$ is independent of $z_0$. Thus multiplying equation
(\ref{general_soln}) with its complex conjugate one obtains the power spectrum:
\begin{equation}
\left|\Psi\right|^2=4\left|H_\omega(z)\right|^2\left|S_\omega^+\right|^2
\left\{ 1+\Re\left( {S_\omega^-\over S_\omega^+}\right) \tan\phi +
{1\over 4}\left|{S_\omega^-\over S_\omega^+}\right|^2\tan^2\phi\right\}
\label{general_power}
\end{equation}
where I have introduced
\begin{equation}
S_\omega^\pm=\int A_{\omega,k}(z_0)e^{\pm\Delta(z_0)}\left| K(z_0)\right|^{-{1\over
2}}\dd z_0
\end{equation}
and
\begin{equation}
H_\omega(z)={ \exp\left[ \int _{Z_t}^z \left| K^2(z')\right|^{1\over 2}\dd z'
\right]
\over
4\left| K^2(z)\right|^{1\over 4} }.
\end{equation}
Treating equation (\ref{general_power}) as a quadratic in $\tan\phi$, it is clear
that
real zeros (true nulls of the spectrum) can exist only if $S_\omega^-/S_\omega^+$ is
real. Thus if we were to replace our delta-function source by a finite-width, 
single-phase
source, the troughs in the spectrum would still be true zeros. Conversely, the
existence of sharp troughs in the measured spectrum cannot be taken as an
indication that the source itself is necessarily highly localised.

\subsubsection{Proposed Line Profile for Low-Frequency Modes}
\label{profile}

A key question in helioseismology is the correct way to determine the true solar
oscillation eigenfrequencies from the measured spectrum of asymmetric, multiply-excited
peaks superimposed on a noisy background. Rast \& Bogdan (1997) suggested that a
correction to
the peak location might be obtained by using the location of the nearest trough
while Rhodes et al. (1997) show a comparison of an asymmetric {\it p}\/-mode
line profile to an asymmetric Lorentzian function.
The asymptotic spectrum (\ref{spec1}) suggests
a possible physically-motivated profile for use in fitting asymmetric peaks
in the data. For a given peak one takes
\begin{equation}
\phi={\pi\over\delta}(\nu-\nu_0)+{\pi\over 2}+i\alpha_1
\end{equation}
where $\delta$ is a typical inter-peak spacing and the damping constant $\alpha_1$
has been introduced to ensure that the amplitude is everywhere finite. Then the
peak profile will be given by
\begin{equation}
|F|^2=\alpha_2+\alpha_3\left|1+\left(\alpha_4+i\alpha_5\right)\tan\left(
{\pi\over\delta}(\nu-\nu_0)+{\pi\over 2}+i\alpha_1\right)\right|^2
\end{equation}
where $\alpha_1$-$\alpha_5$ and $\delta$ are real parameters. The additive noise
background has been expressed as a constant $\alpha_2$, although in practice a power law might
be preferred. 
This can be written
\begin{eqnarray}
|F|^2&=&\alpha_2+\alpha_3\times\cr
&&\!\!\!\!\!\!\!\!\!\!\!\!\!\!\!\!\!\!{
\left\{(\alpha_4\cos x-\sin x)\cosh\alpha_1-\alpha_5\sin x\sinh\alpha_1\right\}^2
+
\left\{(\alpha_4\sin x+\cos x)\sinh\alpha_1+\alpha_5\cos x\cosh\alpha_1\right\}^2
\over
\sin^2x\cosh^2\alpha_1+\cos^2x\sinh^2\alpha_1}\cr
\label{profile1}
\end{eqnarray}
where
\begin{equation}
x\equiv {\pi\over\delta}(\nu-\nu_0).
\end{equation}

One could fit the peak directly to (\ref{profile1}). Alternatively
one notes that
close to the peak we have $\tan\phi\approx -1/(\phi-\pi/2)$ so
\begin{equation}
|F|^2\approx \alpha_2+\alpha_3{
(x-\alpha_4)^2+(\alpha_1-\alpha_5)^2
\over
x^2+\alpha_1^2}
\label{profile2}
\end{equation}
Equation (\ref{profile2}) 
produces only a single asymmetric peak while equation (\ref{profile1}) includes
the influence of nearby peaks directly due to the periodicity of the $\tan$ function.
Thus it is clear by inspection that equation (\ref{profile2}) cannot be used to
fit the parameter $\delta$ independently of the other parameters. 

An important limit of equation (\ref{profile2}) is the case $\alpha_5=0$ which
yields
\begin{equation}
|F|^2\approx \alpha_2+\alpha_3 +{\alpha_4^2\alpha_3\over x^2+\alpha_1^2}
\left(1- {2x\over\alpha_4}\right)
\end{equation}
from which one can see that allowing $\alpha_4$ to tend to infinity while
keeping $\alpha_4^2\alpha_3$ constant yields a symmetric Lorentzian peak. Thus 
the asymmetric forms include symmetric peaks as a special case.

The actual use of either (\ref{profile1}) or (\ref{profile2}) to fit solar
data will be the subject of a future investigation. However, in order to
illustrate that such a procedure is feasible, figure \ref{fit} shows some
preliminary results of a fit of profile (\ref{profile1}) to one month
of low degree GONG data. 
Even from a single month of data, it is clear that the peaks show a negative
asymmetry (more power in the low-frequency wing of the peak than in the high-frequency
wing) and that the asymmetric profile produces a
substantially better fit to the data than does a symmetric Lorentzian.
A full discussion of the fitting procedure, error analysis, and details
of the numerical results will be presented in a future publication, and
will be based on analysis of much longer observational runs from both
GONG and the SOI-MDI instrument on SOHO.

\subsection{The High-Frequency Case}
\label{case2}

In the Sun, oscillations are seen well above the photospheric cutoff frequency. These
have been interpreted (\markcite{Ketal90,KL91} Kumar et al. 1990, Kumar \& Lu 1991) as pseudomodes (interference fringes)
caused by the interference of a wave propagating up from the source with a wave 
propagating downwards from the source and reflected back upwards at its lower 
turning point. The analysis developed in \S\ref{case1} is easily extended to this
case. We now have only three regions to consider: a lower evanescent region, a 
central propagating region below the source and an upper propagating region above the
source. The WKB approximations to the solutions in these three regions are
\begin{eqnarray}
\rm{I:~~~}\Psi\sim& {1\over \left|K^2\right|^{1\over 4}} 
                   A\exp\left[ -\int_{z_t}^z \left| K^2(z')\right|^{1\over 2}\dd z'\right],
                   & z\gg z_t \cr
\rm{II:~~}\Psi\sim&{1\over K^{1\over 2}}\left\{
         B\exp\left[ i\int_{Z_t}^z K(z')\dd z'\right]+C\exp\left[-i\int_{Z_t}^z K(z')\dd
         z'\right] \right\},
         & z_t\gg z >  z_0\cr  
\rm{III:~~}\Psi\sim&{1\over K^{1\over 2}} 
                   D\exp\left[ -i\int_{z_0}^z  K(z')\dd z'\right].
                   &z<z_0
\label{WKB2}
\end{eqnarray}

Using the same matching condition at $z_t$ and jump conditions at $z_0$ leads to
the amplitude spectrum in the upper region:
\begin{equation}
D={1\over 2}(ia_0-a_1)-{1\over 2}(ia_0+a_1)e^{-i(2\tilde\phi+\pi/2)}
\label{spec2}
\end{equation}
where
\begin{equation}
\tilde\phi\equiv\int_{z_0}^{z_t}K \dd z.
\end{equation}
The high-frequency spectrum is thus purely sinusoidal and has no modes or divergences.
However the location of the minima and maxima, their separation, and their visibility are
dependent on both the type and location of the source.

\subsection{The Intermediate-Frequency Case}
\label{case3}

Formally we can consider a third case in which the upper turning point lies well above
the source. In practice this case probably does not accurately describe the properties of
any actual mode occurring in the Sun, so I include it only for completeness and note only
the result for the amplitude spectrum:
\begin{equation}
F={1\over 2\cos\phi}\left[ 
a_0\sin\left(\phi-\theta+{\pi\over 4}\right)-a_1\cos\left(\phi-\theta+{\pi\over 4}\right )
\right]
\label{spec3}
\end{equation}
where $\phi$ is given by equation (\ref{phi}) and 
\begin{equation}
\theta\equiv\int_{Z_t}^{z_0} K\dd z.
\end{equation}
The condition for the eigenmodes is the same as in the low-frequency case [eq. \ref{modes1}]. 
However the absence of
the small factor $\exp-2\Delta$ means that the spectral peaks are much more symmetrical
in this case.

\section{Application to a Simple Model}
\label{results}

Asymptotic descriptions of the properties of solar oscillations have generally proved
useful in helioseismology as guides to the relationship between properties
of the spectrum and properties of the model. Only in limited cases can the actual 
asymptotic expressions themselves be used for direct numerical comparison between
observation and theory. In the present case we have found that the asymptotic analysis
points to the phase-function difference between the modes and the troughs in the
solar spectrum as being a useful diagnostic for the source of the oscillations. I
now apply this to a simple model.

I consider a model consisting of an adiabatic polytropic layer characterized by 
a polytropic index $m$, truncated at some depth $z_{\rm s}$
corresponding to the surface of the Sun, and overlain by an isothermal atmosphere
characterized by a density scale-height, $H_-$. A parameter $\beta$ is defined as the ratio
of the temperature immediately above the surface to that immediately below. Thus
for $z>z_{\rm s}$ we have
\begin{equation}
c_+^2={gz\over m} \quad H_+={z\over m} \quad 
\omega_{c+}^2={mg\over 4z}\left( 1+{2\over m}\right) \hbox{~~and~~} N_+^2=0
\end{equation}
while for $z<z_{\rm s}$ we have
\begin{equation}
c_-^2=\beta{g z_{\rm s}\over m} \quad H_-=\beta{z_{\rm s}\over m+1} \quad
\omega_{c-}^2={1\over \beta}{g(m+1)^2\over 4mz_{\rm s}} \hbox{~~and~~}
N^2_-={g\over\beta z_{\rm s}}.
\end{equation}

Figure \ref{fig001} shows asymptotic power spectra for an $m=3$ polytrope in which
a pure monopole source is located 90km below the photosphere. The peaks
show the same sign of asymmetry as seen in the solar doppler observations. The asymmetry
decreases with increasing horizontal wavenumber and increasing frequency.
Also shown is
the effective difference between the phase function describing the peaks and the 
phase function for the troughs. Evidently the difference is close to being a function of
frequency alone, even towards the higher frequencies where the approximate expression
for the difference 
[eq. \ref{zeros1}] is invalid. 

Figure \ref{fig002} shows the spectrum for the same case as in figure \ref{fig001}
however I now show the result for a single wavenumber but extended over a large frequency range
so as to include all three asymptotic regimes. The crossovers between the three asymptotic
regimes are quite smooth despite the fact that the asymptotic analysis formally breaks down
at those points. The asymmetry is again strongest at low frequency and decreases as the 
frequency is increased. However there is no actual reversal of the sign of asymmetry
for this particular source depth.
In the high-frequency range the troughs are still
present but there are no eigenmode peaks. When the monopole source is replaced
by a dipole the locations of the eigenmodes are, of course, unaffected but the
sign of the line asymmetries is reversed. For the high-frequency modes there is a 
phase shift so that the locations of maxima and minima in the spectrum are interchanged.

\subsection{Comparison of Asymptotic and Numerical Results}

For comparison between the asymptotic and numerical results, I
solve equation (\ref{general_eqn}) with right-hand-side given by 
equation (\ref{local}) using a shooting method. Specifically one starts from
a location well above both the upper turning point and the source with the initial condition
$\Psi'-|K|\Psi=0$ and integrates down to the source at $z_0$.
Call the values of the wave function and its derivative there $\Psi_-$
and $\Psi'_-$. Similarly one can integrate from below the lower turning point up
through the cavity to $z_0$, starting with the boundary condition $\Psi'+|K|\Psi=0$,
to obtain $\Psi_+$ and $\Psi_+'$ immediately below the source. Then the jump
conditions [eqs. \ref{jump1}, \ref{jump2}] imply that the actual solutions above and below
the source are given by $A_-\Psi_-$ and $A_+\Psi_+$ where the coefficients
$A_-$ and $A_+$ must satisfy 
\begin{equation}
A_+\Psi'_+-A_-\Psi'_-=a_0|K(z_0)|^{1\over 2}
\end{equation}
and
\begin{equation}
A_+\Psi_+-A_-\Psi_-=a_1|K(z_0)|^{-{1\over 2}}
\end{equation}
giving
\begin{equation}
A_-= {{ a_0|K(z_0)|^{1\over 2}\Psi_+ - a_1|K(z_0)|^{-{1\over 2}}\Psi'_+ }
     \over
     {\Psi_+'\Psi_- -\Psi_-'\Psi_+ }}.
\end{equation}  
The condition for an eigenmode is therefore that the denominator vanishes, while
the condition for a zero in the power is that the numerator vanishes.
Without loss of generality we may assume that $\Psi_+$ and $\Psi_-$ are real. If we
consider only cases where $a_0$ and $a_1$ are in phase, or cases in which
either $a_0$ or $a_1$ vanishes, we see that zeros of $A_-$ can occur and that their
location is independent of $\Psi_-$.
Thus {\it the location of the nulls in the spectrum is independent of
the atmospheric structure above the source}\/. The explanation for this result
is quite simple - since the nulls correspond to
frequencies where the amplitude in the superficial layers is identically zero
their location cannot depend on the structure of those layers. In the more
general case in which $a_0$ and $a_1$ are complex the minima will no longer be true zeros and their location will no
longer be independent of $\Psi_-$. Instead they will be given by the condition
$\dd |A_-|^2/\dd \omega=0$ which in general will depend also on $\Psi_-$. 
Nevertheless, the minima will still be locations in which the amplitude in the
region above the source is small and their location will therefore still be
less sensitive to the structure of the superficial layers than the locations
of the eigenmodes. This result is illustrated in figure \ref{fig004}
which shows a single peak and trough in the case of a monopole and dipole
source acting in phase-quadrature and for two different values of the
photospheric temperature-jump parameter. Evidently the trough remains almost fixed
while the eigenfrequency varies.

If we return to the asymptotic expressions for the eigenfrequencies and
nulls [eqs. \ref{modes1} and \ref{zeros1}] we see that these expressions
imply that the mode frequencies depend only on the structure below the
upper turning point while the nulls depend only on the structure below the source.
However, while the former statement is only an approximation, because the
eigenfunctions have non-zero amplitude outside the turning points, the latter statement
is exact for those cases where the troughs are true nulls. 

Figure \ref{fig003} shows a comparison of numerical and asymptotic estimates of the
power spectrum. While the spectra are qualitatively the same, quantitatively the
separation between the peaks and troughs for the numerical calculation is much
smaller than that predicted by the asymptotics. This suggests that while the asymptotics
may be a useful guide as to the correct interpretation of the line asymmetries, 
it would be wrong to use an asymptotic expression such as equation (\ref{dalpha}) 
in a quantitative comparison with the data.

\section{Discussion}

Typically, helioseismology has concentrated on analysis of the eigenmode frequencies
of the Sun and relatively little attention has been paid to the rest of the spectrum.
With the advent of the SOI/MDI instrument on SOHO and the GONG network, noise levels
in measured spectra are now sufficiently low that it is not only possible to make useful
measurements between the mode peaks but {\it necessary}\/ to do so since it is
increasingly clear that the line asymmetry can produce systematic
errors in the determination of {\it p}\/-mode frequencies if it is not accounted for
properly. 

A puzzle in the data has been the observation originally due to Duvall et al (1993),
and now apparently confirmed from SOI-MDI (Nigam et al. 1998), that
the sign of the measured asymmetry is oppositely directed in doppler and
intensity measurements. The doppler measurements imply a dominant monopole source
(a negative asymmetry in the Abrams \& Kumar terminology) while the intensity
measurements would suggest a dipole source. Rast \& Bogdan (1997) show that, 
regardless of the form of the source function, one would expect to see the 
same sign of asymmetry with observations made in different variables, since
the locations of the peaks and troughs are the same for all variables. 
Nigam et al. have shown that the phenomenon can be explained if the measured
signal in either doppler or intensity is contaminated by a noise source which is
correlated with the source, for example if the measured signal contains a
contribution coming directly from the source itself. The question is then raised as
to which of the doppler measurement and the intensity measurement actually
constitutes a better measurement of the true acoustic spectrum. One might argue
that the doppler signal is probably less contaminated since it has a lower overall
background level, but it is certainly possible that both the intensity and
doppler measurements contain some leakage from the source, with an amplitude
and phase which are not yet precisely known. Evidently the possibility that the
location of the spectral minima is shifted due to such leakage must complicate the
interpretation of the line-asymmetry in terms of source location and type,
and further work will be required to determine whether the combined doppler and
intensity measurements can be used to determine the true acoustic spectrum in
a model-independent fashion.

As originally pointed out by Rast \& Bogdan, the asymmetry is best approached by consideration of the minima
in the spectrum, whose location is determined by the depth and type of the 
source of the oscillations. In this work I have determined an asymptotic expression
for the frequency spectrum, from which it is possible to determine the locations
of the minima. The sign of the line asymmetry is determined by whether the
nearest minimum lies above or below the eigenfrequency. The former case corresponds
(at low frequency) to a pure monopole source and the latter to a pure dipole. The
amount of asymmetry depends of the location of the source. The higher the source is
placed above the upper turning point of the oscillation the closer the trough will
lie to the eigenmode and therefore the greater the asymmetry. Thus the extent of the 
asymmetry should be a useful diagnostic of the properties of the turbulent
superadiabatic layer in which the source of the solar \pmodes~is believed to lie.

Numerical results confirm the qualitative properties of the asymptotic results.
Also for the case of a pure monopole or dipole, or where the two components
of the source are in phase, the exact locations of the nulls (including those
above the acoustic cutoff frequency) can be determined by
solving numerically the oscillation equations subject to the boundary condition
\begin{equation}
{\dd\Psi\over \dd z}- {a_0\over a_1} |K| \Psi=0
\label{bc1}
\end{equation}
at $z=z_0$. This should be relatively straightforward to implement in most existing
oscillation codes. One can thus determine numerically
the phase-function difference $\alpha_t-\alpha_m$ which is predicted by the
asymptotic analysis to be a key diagnostic of the structure and location of the source.

Since the nulls are independent of the superficial structure the asymmetry could, in principle,
provide a 
tool for differentiating between superficial and more-deeply seated changes in the Sun.
In particular, solar-cycle variations occurring above the source location, in the lower 
atmosphere for example, would change the mode frequencies but not the trough
frequencies and could thus be visible as changes in the asymmetry, while changes
below the source would change the frequency of both the mode and the trough.
In practice such changes are likely to be extremely difficult
to observe, not least because the asymmetry is most evident for lower frequency
modes for which the solar-cycle dependent changes are small. 

We have thus reached the stage where it is possible to envisage a program for the 
re-analysis of helioseismic data to make full use of the information contained
in the \pmode~line profiles. In particular it would be highly desirable to 
re-analyse the spectral data using parameterised line-profiles such as equation
(\ref{profile1}) or equation (\ref{profile2}) in order to obtain improved estimates
of the eigenfrequencies simultaneously with estimates for the spacing between
the maxima and minima. From these we could obtain an observational $\alpha_t-\alpha_m$
curve which could be calibrated against theory using nulls calculated from an
eigenvalue calculation with boundary condition (\ref{bc1}). Ultimately one would hope
that the results of such a calibration could be brought into line with predictions made
from numerical simulations of the solar surface layers ({\it
e.g.}\/\markcite{SN89,SN91,Netal96} Stein \& Nordlund 1989, Stein \& Nordlund 1991,
Nordlund, Stein \& Brandenburg 1996) to 
provide a solidly grounded understanding of this highly complex region of the Sun.

\acknowledgments

The National Center for Atmospheric Research is sponsored 
by the National Science Foundation. This research was supported by
SOI/MDI NASA GRANT NAG5-3077. This work utilizes data obtained by the Global Oscillation
Network Group (GONG) project, managed by the National Solar Observatory, a
Division of the National Optical Astronomy Observatories, which is operated by
AURA, Inc. under a cooperative agreement with the National Science
Foundation.
The data were acquired by instruments operated by the
Big Bear Solar Observatory,
High Altitude Observatory,
Learmonth Solar Observatory,
Udaipur Solar Observatory,
Instituto de Astrofisico de Canarias,
and
Cerro Tololo Interamerican Observatory.
I would like to thank Mark Rast, Tom Bogdan, and Pawan Kumar for a number of
enlightening discussions and helpful suggestions.

\clearpage

\begin{figure}
\caption{
A schematic of the illustrative model used in \S\ref{illus}.
}
\label{fig1}
\end{figure}

\begin{figure}
\caption{
Power (solid line) and energy (dotted line) spectra for the illustrative model
for three different sets of parameter values.
}
\label{fig2}
\end{figure}

\begin{figure}
\caption{
A schematic illustration of the geometry for the low-frequency regime
analysed in \S\ref{case1}.
}
\label{fig4}
\end{figure}


\begin{figure}
\caption{
Preliminary results of a least-squares fit of equation (\ref{profile1}) to 
$m$-averaged power spectra obtained from one month of GONG data. Each panel
is an average over all \pmode~peaks in the range $0\le l\le 10$ in the frequency
bin indicated, with the mode peaks aligned using the most recent GONG 
tables of frequencies and splitting coefficients. The dashed line shows the fit
to the asymmetric profile (\ref{profile1}) and the dotted line is a fit to
a Lorentzian profile with a constant background.
}
\label{fit}
\end{figure}

\begin{figure}
\caption{
Asymptotic power spectra resulting from driving an $m=3$ polytrope with a
monopole source located 90km below the photosphere. Results are shown for
three different values of the mode degree $l$, defined by $l=kR_\odot$. 
The lower panel shows the difference in phase function between the trough
and the mode for each trough-mode pair appearing, using the symbols
($\diamond$) for $l=100$, ($+$) for $l=200$ and ($*$) for $l=300$.
}
\label{fig001}
\end{figure}

\begin{figure}
\caption{
Asymptotic power spectra over a broad frequency range for the same polytropic index
and source depth as used in figure \ref{fig001} for $l=200$. The solid line is for a pure monopole
source and the dotted line is for a pure dipole source. The crossovers between
the three asymptotic regimes are marked by vertical dashed lines.
}
\label{fig002}
\end{figure}

\begin{figure}
\caption{
Numerical power spectra for a single mode with $l=300$ for the two cases
$\beta=1.5$ (dotted line) and $\beta=10^{-3}$ (solid line) for a
source characterised by $a_0/a_1=-10i$ (and with other parameter values as
in figure \ref{fig001}). 
}
\label{fig004}
\end{figure}

\begin{figure}
\caption{
Comparison of numerical (dotted line) and asymptotic (solid line) low-frequency
spectra for $l=200$. The lower panel shows the frequency difference between a trough
and its neighbouring peak for each pair in the calculation.
}
\label{fig003}
\end{figure}

\end{document}